\documentclass[11pt,dvips]{article}

\usepackage{picinpar}
\usepackage{wrapfig}
\usepackage{floatflt}
\usepackage{psfig}
\makeatletter
\renewcommand{\section}{\@startsection{section}{1}{0in}
	{0.4\baselineskip}{0.1\baselineskip}{\Large\bf}}
\renewcommand{\subsection}{\@startsection{subsection}{2}{0in}
	{0.25\baselineskip}{-\baselineskip}{\large\bf}}
\renewcommand{\subsubsection}{\@startsection{subsubsection}{3}{0in}
	{0.1\baselineskip}{-\baselineskip}{\normalsize\bf}}

\makeatother
\newcommand{\vol}[2]{$\,$\rm #1\rm , #2.}     
\def\teq#1{$\, #1\,$}
{\catcode`\@=11                                                  
\gdef\SchlangeUnter#1#2{\lower2pt\vbox{\baselineskip 0pt\lineskip0pt    
\ialign{$\m@th#1\hfil##\hfil$\crcr#2\crcr\sim\crcr}}}}           
\def\gtrsim{\mathrel{\mathpalette\SchlangeUnter>}}               
\def\lesssim{\mathrel{\mathpalette\SchlangeUnter<}}   
\def\apj{Astrophys. J.}                         
\def\apjl{Astrophys. J.}                        
\def\nat{Nature}                                

\begin{document}

%
\thispagestyle{myheadings}
%
\markright{OG.2.4.17}
\begin{center}
%
{\LARGE \bf Photon Attenuation and Pair Creation\\[4pt]
            in Highly-Magnetized Pulsars}
\end{center}

\begin{center}
%
%
{\bf Matthew G. Baring$^{1,2}$ and Alice K. Harding$^1$\\}
{\it $^{1}$Laboratory for High Energy Astrophysics, NASA/GSFC, Greenbelt, MD 20771, USA\\
$^{2}$Universities Space Research Association}
\end{center}

\begin{center}
{\large \bf Abstract\\}
\end{center}
\vspace{-0.5ex}
%
%
Developments over the last couple of years have supported the
interpretation that anomalous X-ray pulsars (AXPs) and soft gamma
repeaters (SGRs) possess unusually high magnetic fields, and
furthermore may represent a class or classes of neutron stars distinct
from the population of conventional radio pulsars.  We have recently
suggested that such a dichotomization of the pulsar population may
naturally arise due to the inherently different conditions in
subcritical and supercritical magnetic fields.  In this paper, we
summarize, within the polar gap model, expectations for observable
properties of highly magnetized pulsars, conventional or anomalous.
This includes a discussion of the potential suppression of pair
production and cascade generation in very strong fields by photon
splitting and by threshold pair creation, which might explain radio
quiescence in AXPs and SGRs.  X-ray and hard gamma-ray spectral
properties and trends are identified, with a view to establishing goals
for future high energy experimental programs.
%

\vspace{1ex}

\section{Introduction}
\label{intro.sec}
The study of pulsars with unusually high magnetic fields, namely
\teq{B\gtrsim 10^{13}}G, has recently become of great interest in the
astronomical community, due both to the rapid increase in observational
data indicating such high fields, and also to the fascinating physics
that might arise in their environs.  Apart from a handful of
conventional radio pulsars such as PSR 1509-58 with dipole spin-down
field estimates in the range \teq{10^{13}}G\teq{\lesssim B_0\lesssim
3\times 10^{13}}G, there is the growing body of anomalous X-ray pulsars
(AXPs) and soft gamma repeaters (SGRs) perhaps with much larger
fields.  Such sources are currently being touted as candidate {\it
magnetars} (e.g.  Thompson \& Duncan 1993), a class of neutron stars
with fields in excess of \teq{10^{14}}G.

The AXPs are a group of six or seven pulsating X-ray sources with
periods around 6-12 seconds, which are anomalous in comparison with
average characteristics of known accreting X-ray pulsars.  They are
bright, steady X-ray sources having luminosities \teq{L_X \sim
10^{35}\,\rm erg\; s^{-1}}, they show no sign of any companion, are
steadily spinning down, and have ages \teq{\tau \lesssim 10^5} years.
Those which have measured \teq{\dot P} (e.g. Mereghetti \& Stella 1995;
Gotthelf \& Vasisht 1998) have derived dipole spin-down magnetic fields
between \teq{10^{14}} and \teq{10^{15}} Gauss.  The SGRs, so-named
because of repeated transient $\gamma$-ray burst activity, are another
type of high-energy source that has recently joined this group of
possible magnetars.  There are four known SGR sources, one newly
discovered (Kouveliotou, et al. 1998b), and two (SGRs 1806-20 and
0526-66) being associated with young (\teq{\tau < 10^5} yr) supernova
remnants.  Recently, 7.47s and 5.16s pulsations have been discovered
(Kouveliotou, et al. 1998a,c; Hurley et al. 1998) in the quiescent
X-ray emission of SGRs 1806-20 and 1900+14, respectively, with SGR
1900+14 exhibiting a 5.15s period in a $\gamma$-ray burst (Hurley et
al. 1999), much like the original and canonical 5th March 1979 event
from SGR 0526-66.

In the pulsar \teq{P-\dot P} diagram, both AXPs and SGRs live in a
separate region above the detected radio pulsars:  no radio pulsars
have inferred fields above \teq{\sim 10^{14}} Gauss, even
though known selection effects do not {\it a priori} prevent their
detection.  This motivated Baring \& Harding (1998) to propose an
explanation for the absence of radio pulsars of such high
magnetization.  We identified a potentially strong suppression of pair
creation, \teq{\gamma\to e^{\pm}}, in fields above \teq{\sim 10^{13}}
Gauss, ultimately through the action of the exotic QED process of
photon splitting \teq{\gamma\to\gamma\gamma}.  In this paper, we
present results of a more detailed exploration of this issue, namely a
determination of pair yields from simulations of pair cascades.  Our
computations indicate that photon splitting is generally only
marginally effective at reducing the number of pairs when only one
polarization mode (\teq{\perp\to\parallel\parallel}) of splitting operates
(according to kinematic selections rules derived by Adler 1971), due to
significant pair production by photons of \teq{\parallel}
polarization.  In this case, ground state (threshold) pair creation
significantly reduces the total number of pairs.  If three polarization
modes of splitting operate, then pair suppression is dramatic and the
contention of Baring \& Harding (1998) is borne out.

Throughout this paper, the standard convention for the
labelling of the photon polarizations will be adopted, namely that
\teq{\parallel} refers to the state with the photon's {\it electric}
field vector parallel to the plane containing the magnetic field and
the photon's momentum vector, while \teq{\perp} denotes the photon's
electric field vector being normal to this plane.

\section{Pair Suppression and Radio Quiescence?}
 \label{results.sec}
A premise of standard polar cap models for radio pulsars is that a
plentiful supply of pairs is a prerequisite for coherent radio emission
at observable levels.   By extension, an immediate consequence of the
significant suppression of pair creation by splitting in pulsars (and
other effects mentioned just below) is that detectable radio fluxes
should be strongly inhibited.  Baring \& Harding (1998) determined an
approximate criterion for the {\it boundary of radio quiescence} in the
\teq{P-\dot P} diagram based on a comparison of attenuation properties
for \teq{\gamma\to e^{\pm}} and \teq{\gamma\to\gamma\gamma} in general
relativistic neutron star magnetospheres.  They found that for
\teq{\dot{P}} above \teq{\dot{P}\approx 7.9\times 10^{-13}\,
(P/1\,\hbox{sec})^{-11/15}}, photon splitting by the
\teq{\perp\to\parallel\parallel} mode should dominate pair creation by
\teq{\perp}-polarized photons, corresponding to fields in the range
\teq{3\times 10^{13}}G\teq{\lesssim B_0\lesssim 8\times 10^{13}}G,
thereby defining the lower extent of a region of radio quiet pulsars.
Underpinning this is the fact that while \teq{\parallel}-polarized
photons can still produce pairs, they are in relative paucity in
multi-generational cascades due to the predominance of \teq{\perp}
photons (\teq{\gtrsim 75\%}) in the continuum emission processes of
curvature and synchrotron (or cyclotron) radiation and resonant Compton
upscattering (Baring and Harding 1999; hereafter BH99).  Clearly, the
suppression of pair creation by splitting {\it is partial, not total}.

The robustness of this putative boundary for radio quiescence, which is
computed specifically for photon origin near the stellar surface, can
be assessed with a Monte Carlo cascade calculation.  We have developed
(BH99) a simulation that extends the work of Harding, Baring and
Gonthier (1997).   It follows photons from a point above the stellar
surface, along trajectories in curved spacetime, permitting them to
split or create pairs, and then computing curvature and
synchrotron/cyclotron radiation products of these pairs.  Subsequent
attenuation of these photons is determined, as are the next generation
of pairs and their emission; a complete computation of photon-initiated
pair cascades is thus obtained.  The principal missing ingredient in
this simulation is a consistent determination of the point of emission
of the primary photon in conjunction with an acceleration region.

In Figure~1, we depict {\it pair yields}, the number of pairs produced
per primary photon, for an input primary photon spectrum typical of
that for curvature radiation from uncooled monoenergetic electrons.
The photons assume a power-law of index \teq{\alpha\sim 1.6}, cutoff of
at various energies \teq{\omega_{\rm max}m_ec^2}, as indicated.  The
yields are functions of the surface field strength \teq{B_0}, in units
of the quantum critical field \teq{B_{\rm cr}=4.413\times
10^{13}}Gauss.  While the left panel of the Figure explores variations
with the maximum photon energy, and differences obtained when photon
splitting is present or is artificially suppressed, the right panel
illustrates the effect of changing the polar cap size \teq{\Theta}
(assumed to be in degrees here), or pulsar period \teq{P\approx
0.69\Theta^{-2}}.

The first obvious feature is a drop in the pair yield at surface fields
of \teq{B_0\sim 3\times 10^{12}}Gauss.  When the local field is
\teq{B\gtrsim 6\times 10^{12}}Gauss, pair creation occurs predominantly
in the lowest accessible Landau state configuration (Harding \&
Daugherty 1983).  In such fields, \teq{\perp}-photons produce pairs no
higher than the first Landau level so that subsequent cyclotron photons
are mostly below pair threshold.  Since \teq{\parallel}-photons leave
pairs in the ground (0,0) state, they spawn no cyclotron/synchrotron
emission, preventing any further pair generations.  Hence pair
cascading is strongly inhibited for such local fields, and this appears
as a decline in the pair yields for high \teq{\omega_{\rm max}} cases
with large polar cap sizes.   Next, the left panel indicates that, in
the case of one splitting mode, the number of pairs produced per
primary photon saturates at high fields to constant values that depend
on the polar cap colatitude \teq{\Theta} and \teq{\omega_{\rm max}}.
This arises principally because the attenuation rates for both pair
creation and photon splitting saturate in ultra-quantum fields.  For
really low \teq{B_0}, the pair yield drops off, due to the associated
decline in the pair creation rate.  In the right hand panel, the
correlation of pair yield with \teq{\Theta} just reflects the reduction
in the pair creation rate with larger radii of field curvature.  These
last two trends putatively couple to the conventional death line for
radio pulsars.

\begin{figure}[t]
\centerline{\hskip 1.7truecm
             \psfig{figure=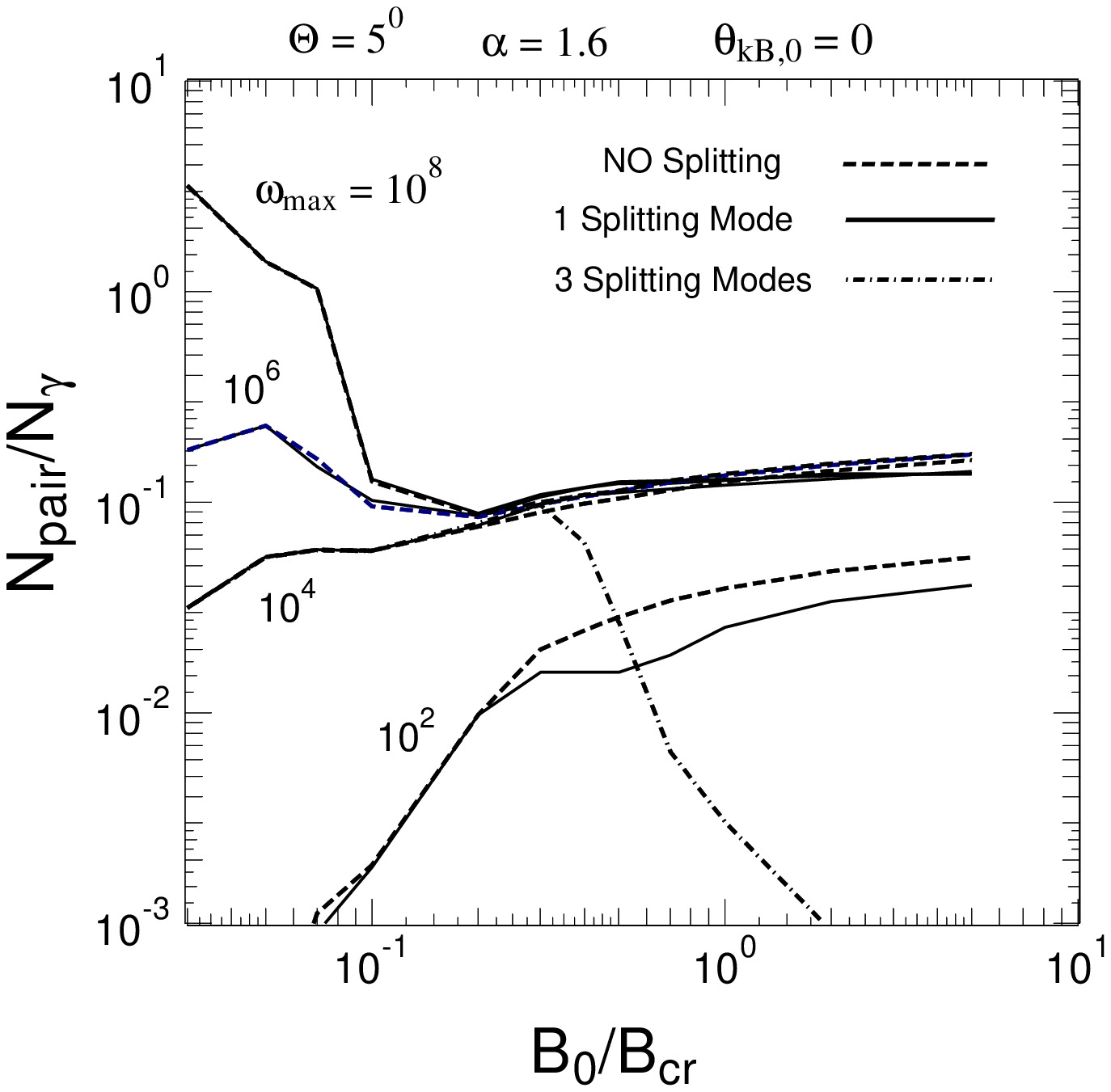,width=11.7cm}\hskip -2.5truecm
              \psfig{figure=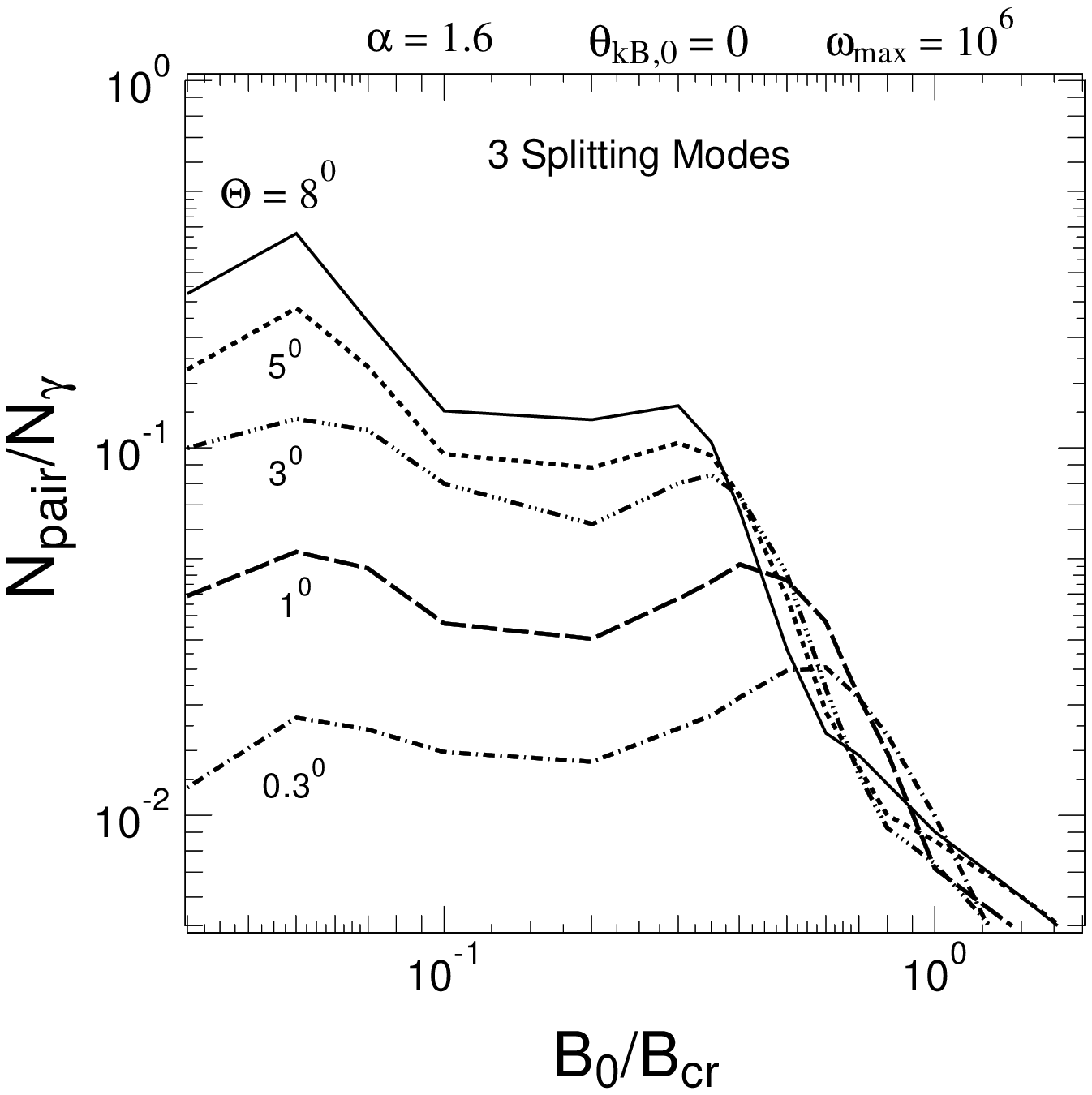,width=11.7cm}}
\vspace{-10pt}
\caption{
The cascade pair yield (number of pairs per injected photon) as a
function of surface magnetic field strength, $B_0$, in units of the
critical field, $B_{\rm cr}$.  {\it (Left Panel)} Dependence of the
pair yield on the maximum primary photon energy $\omega_{\rm max}$ (in
units of $m_ec^2$) is displayed.  Different line types refer to the
cases where no splitting, only one mode, or three modes of splitting are
allowed (see text).
{\it (Right Panel)}
Here the pair yields are exhibited for varying primary photon injection
colatitudes, $\Theta$, for the case where three photon splitting modes
operate.  Strong suppression of pair creation is afforded in these
cases for high $B_0$.
 }
   \label{fig:icrc99psr_f1}
\end{figure}

Inspection of the left hand panel of Figure~1 immediately reveals that
there is miniscule reduction in the pair yield by the action of one
polarization mode (\teq{\perp\to\parallel\parallel}) of splitting in
all but the lowest \teq{\omega_{\rm max}} case.  The reason for this is
twofold: (i) the onset of pair creation in the ground state at lower
fields has already inhibited cascading to the point that there is
little more that splitting can do, and (ii) the splitting
\teq{\perp\to\parallel\parallel} actually creates more photons than it
destroys, and all of these are available for pair creation at a
somewhat lower threshold (than for \teq{\perp} photons).   Hence,
splitting actually both hinders and helps pair creation, the two
tendencies effectively negating each other.  This situation changes
when all three splitting modes (\teq{\perp\to\parallel\parallel},
\teq{\perp\to\perp\perp}, \teq{\parallel\to\perp\parallel}) allowed by
CP invariance are permitted to operate.  Then pair creation is rapidly
quenched with increasing \teq{B_0}, an effect whose onset \teq{B_0}
declines with larger \teq{\Theta}, and matches well the predictions of
Baring \& Harding (1998).  It is presently unclear how many modes of
splitting operate in supercritical fields: assessment of this would be
based on the extremely-involved calculation of higher-order quantum
electrodynamical dispersive effects in the magnetized vacuum.  Details
of these results and discussion of these and other issues can be found
in Baring and Harding (1999).

It is also salient to briefly address expectations for the X-ray and
gamma-ray properties of highly-magnetized pulsars in their quiescent
states.  The radio quiescence boundary is transparent in such
observational bands since pulsar X-ray and gamma-ray fluxes are not
expected to be strongly correlated with the number of pairs.  This is
because cascading mechanisms merely redistribute the emission between
the gamma-ray and X-ray bands without severe diminution of the overall
luminosity.  Since high field pulsars have cyclotron energies in the
hard X-ray/soft gamma-ray range, soft and hard X-ray experiments will
probe the cyclotronic and sub-cyclotronic structure in the spectra of
high field pulsars.  One expects these sources to exhibit spectral
bumps and breaks near the cyclotron fundamental.  Furthermore, given a
handful of sources, experiments such as Integral can search for trends
with spin-down field.  For example, as pair suppression ensues above
\teq{\sim 3\times 10^{12}}G, a decline on the number of pair
generations and loss of the steeper synchrotron component to spectra
are expected, corresponding to flatter spectra (case in point,
PSR1509-58).  At the same time, the cyclotron fundamental moves up in
energy.  An observed coupling between such effects would strongly argue
in favor of the polar cap model for high energy pulsar emission.  In
addition, hard gamma-ray missions such as GLAST can explore the
expected anti-correlation between the maximum observable photon energy
(due to attenuation by pair production and photon splitting) and the
surface field \teq{B_0}, a prediction of the polar cap model that is
difficult to replicate in outer gap scenarios.  It is important to note
that while these properties accommodate sources like PSR 1509-58, they
are at odds with what little is known of the AXPs and SGRs, which
display moderate to steep soft X-ray spectra in quiescence, as
determined by RXTE and ASCA.  It is possible that these steep spectra
are an entirely different component from the curvature/inverse
Compton/synchrotron cascade emission addressed in this paper.

%
\vspace{1ex}
\begin{center}
{\Large\bf References}
\end{center}
Adler, S.~L. 1971, Ann. Phys. \vol{67}{599}
\\
  Baring, M.~G. \& Harding A.~K. 1998, \apjl\vol{507}{L55}
\\
  Baring, M.~G. \& Harding A.~K. 1999, in preparation.
\\
  Gotthelf, E.~V. \& Vasisht, G. 1998, New Astronomy \vol{3}{293}
\\
  Harding A.~K., Baring, M.~G. \& Gonthier, P.~L. 1997, \apj\vol{476}{246} 
  (HBG97)
\\
  Harding, A.~K. \& Daugherty, J.~K. 1983, in Positron Electron Pairs in 
  Astrophysics, eds. \\
 \hbox{\hskip 10pt}   M.~L. Burns, et al.
  (AIP Conf. Proc. 101: New York), p.~194.
\\
  Hurley, K. et al. 1998, IAU Circ. No. 7001.
\\
  Hurley, K. et al. 1999, \nat\vol{397}{41}
\\
  Kouveliotou, C. et al. 1998a, \nat\vol{393}{235} 
\\
  Kouveliotou, C. et al. 1998b, IAU Circ. No. 6944.
\\
  Kouveliotou, C. et al. 1998c, IAU Circ. No. 7001.
\\
  Mereghetti, S. \& Stella, L. 1995, \apj\vol{442}{L17}
\\
  Thompson, C. \& Duncan, R. C. 1993, \apj\vol{408}{194}

\end{document}